\documentstyle[11pt,epsf,paspconf]{article}
\begin{document}

\title{Clusters of Galaxies and Mass Estimates}

\author{R. Sadat\altaffilmark{1,2}}
\affil{Observatoire astronomique de Strasbourg, 11 rue de l'Universit\'e,
67000 Strasbourg, France}
\affil{C.R.A.A.G., BP 63, Bouzareah, Algiers, Algeria\\
sadat@wirtz.u-strasbg.fr}

\begin{abstract}
This talk is a brief review of the different methods of galaxy cluster mass 
estimation. The determination of galaxy cluster mass 
is
of great importance since it is directly linked to the well- known problem 
of dark matter in the Universe and to the cluster baryon content.
X-ray observations from satellites have enabled a better understanding of the physics
occuring inside clusters, their matter
content as well as a detailed description of their structure.
In addition, the discovery of giant gravitational arcs and the 
lensing properties of clusters of galaxies represent the most 
exciting events in cosmology and have led to many new results on 
mass distribution.
In my talk, I will review some recent results concerning 
the mass determination in clusters of galaxies.   
\end{abstract}
\section{Introduction}

Clusters of galaxies are the most extended gravitationally 
bound systems. They provide an ideal tool for cosmologists
to study the formation and evolution of the structures of the 
Universe. They present strong evidence for the presence of large amounts of 
dark matter. Therefore it is essential to determine in a very accurate way 
their gravitational masses to better constrain the still unknown cosmological 
density parameter ${\Omega}_{0}$.\\ 
Historically, the evidence of the presence of a huge missing mass was derived from 
the application of the standard virial theorem (Zwicky 1933), which is based 
on the assumption that 
mass follows the light distribution, but this assumption has not yet been confirmed. In this 
talk I will show that the total 
cluster mass depends on the relative distribution of the visible and invisible 
components and I will discuss the accuracy of the masses derived under the mass-follows-light 
assumption. \\
Clusters of galaxies, are also strong X-ray emitters. Since the discovery of 
the hot diffuse gas responsible for X-ray emission, astronomers have 
started to use X-ray
observations to constrain cluster masses. 
Methods based on such observations have several advantages compared to optical 
methods. However, it is not yet clear how accurate the standard methods such as the 
hydrostatic ${\beta}$-model are. 
ROSAT observations of the Coma cluster have led to a large fraction of baryons 
in contradiction with the 
standard Big Bang nucleosynthesis predictions. This {\sl baryon catastrophy} has 
several implications 
for cosmology in particular on the value of the density of the Universe, ${\Omega}_{0}$ .\\  
Finally, the detection of gravitational lensing in clusters of galaxies has provided astronomers with the most 
powerfull tool for mapping the mass distribution. The mass estimates using the
lensing method are in general in good agreement with the optically 
derived masses while the X-ray method has systematically
underestimated cluster masses by a factor 2-3.
I will first describe briefly the observational properties of clusters of 
galaxies, then I will review different methods 
which are usually used to estimate their masses and discuss their reliability.\\
In this paper, I will adopt the value of $H_0=50$ km s$^{-1}$ Mpc\\

\section{Observational properties}

\subsection {Optically}
Optically clusters of galaxies appear as large concentration of galaxies in a small volume. 
A typical cluster has several hundred
$\approx{1000}$ of galaxies, which are mainly ellipticals and SOs in irregular clusters. The typical scale radius is 
about 1Mpc.  The distribution of these galaxies has most traditionally been fit by an isothermal 
gravitational sphere which has the approximate analytical form given by King's 
model \\
\begin{equation}
n_{gal}(r) {\propto[1+r/R_{c}]^{-1}},
\end{equation}

\noindent where r is the projected radius and 
$R_{c} \approx{0.5} {\rm Mpc}$ is the core radius in a typical cluster.\\
The radial velocities of the cluster members 
in a well-relaxed cluster are distributed according to a Gaussian 
distribution.\\
\begin{equation}
N(v{_r}) dv{_r} {\propto \exp{(-v{_r}/2{\sigma}{^2})}},
\end{equation}

\noindent where $\sigma{^2}= <(v{_r}-<v>)^2>$ is the line- of- sight velocity 
dispersion\\
\noindent
Merritt (1994), has shown that the mass distribution can be constrained from 
an analysis of the 
shape of radial velocity histograms, but his method requires a 
large number of measured radial velocities. Indeed, a redshift survey of rich 
clusters of galaxies has typically ${\approx 50}$ velocity measurements per 
cluster, making this 
method unusable except in the case of 
the well-studied rich cluster Coma which has ${\approx 600}$ measured radial 
velocities, 
but still, there is the problem of substructures.

\subsection{X-ray emission}

The X-ray emission from clusters of galaxies is mainly due to hot and diffuse intra-cluster gas with 
$T{_x} \approx 10{^7} - {10^8}$ K and a central density of $n_{x}(0)
\approx {10^{-3}} cm^{-3}$
(see the excellent review by Sarazin 1986)\\
This hot intracluster gas is the main baryon component of clusters of galaxies: its mass is several times that 
of the stellar mass $M_{g} \approx 5 - 7 M_{*}$ (David {\it et al.} 1994). It represents a large fraction of the total mass (visible+dark matter) and 
can reach values of {30\%} of the total binding mass (B\"ohringer, 1994).
This gas radiates by thermal bremsstrahlung emission \\

\begin{equation}
{\epsilon}_{\nu}=n{_x}^2 T{_x}{^{-1/2}}\exp{(-h{\nu}/kT{_x})}.
\end{equation}

\noindent For very hot gas the spectrum is dominated by the continuum and the 
only line which is detected in this continuum is the iron 
line. At cooler temperatures however, some heavy 
element emission lines such as O, Si, S, Ar and Ca start to appear.\\
The detection in the X-ray spectra of the iron K-line at 6 kev has shown that the gas has been enriched in 
metals. These metals have been processed into cluster 
galaxies and ejected into the ICM through SN driven winds or outflows, providing evidence of
a non-primordial origin of part of the gas. The typical abundances are  about 
1/3 -1/2 solar (Mushotzky 1996).
What is the quantity of the ejected gas ? And what type of galaxies enriched the ICM? 
All these questions are still open (Arnaud 1994). For the mechanism of metal enrichment of 
the ICM, it is now well accepted that supernovae are reponsible for the injection into the 
ICM of the 
heavy elements processed into stars but we do not yet understand the relative 
importance of both types (Matteuci this school) .\\

\subsection{The Baryon Catastrophy}

Standard Big Bang nucleosynthesis predictions of the primordial abundances place tight 
limits on the present day baryon density in the Universe, 
\[0.04 < \Omega{_b}^{BBN} {\it h^2} < 0.05\] Walker {\it et al.} (1991). This is only a 
small fraction of the critical 
closure density of the Universe.\\
White {\it et al.} (1993) have noted that hot gas in the Coma cluster contributes 
$\sim 15\%$ of the 
total mass within the Abell radius. Thanks to the wide 
field of view and high sensitivity of the ROSAT satellite, it has been possible to 
reliably measure the baryon fraction of the Coma cluster to an even much larger radius 
$\sim 4 {\rm Mpc}$
(Briel, Henry \& B\"ohringer, 1992), where this fraction reaches the value of 
30{\%}.  
If dark matter is distributed similarly to the 
X-ray gas, the conservative value of the gas fraction in Coma cluster 
$f_{b}\sim 15{\%}$ leads to $\Omega_{b} \sim{0.15}$, 
which is ${\sim 3}$ times 
the universal $\Omega{_b}^{BBN}$ value. 
Previous X-ray analyses of galaxy clusters with the Einstein and EXOSAT observatories have already 
found high baryon fractions, but the authors have not emphasized the 
implications of such quantities 
of baryons. More recently, compilations of X-ray cluster data and their analysis 
by White \& Fabian (1995, hereafter WF) and David, Jones \& Forman (1995, hereafter DJF), have led to the same conclusion, showing that 
the problem of baryon overdensity is common in clusters of galaxies. What are 
the 
cosmological implications of this result? The most obvious one is that 
$\Omega_{o}$ is less than one. 
Indeed, one way to reconcile the baryon fraction from cluster analyses 
($f_{b}\sim 15{\%}$) with the 
primordial nucleosynthesis prediction,
is that  $0.26 < \Omega{_o} h^{1/2} < 0.33$. That means that the Universe 
is open. Recent measurements of the primordial deuterium abundance D/H from quasar 
absorption line spectra have produced two different values, a low value 
${\Omega{_b} h_{100}^2}= 6.2 \pm 0.8 10^{-3}$ (Rugers \& Hogan 1996) and a high 
value ${\Omega{_b} h_{100}^2}= 0.024 {\pm 0.006}$ (Tytler, Fan \& Burles 1996). If one 
accepts the higher D/H value and accounts for baryons within clusters of 
galaxies then $\Omega{_o} h^{1/2} {\sim 0.6}$.
The $\Omega{_o} = 1$ universe can 
be rescued if one believes either in a low Hubble constant 
($H_{o} < 40$ Figure 1, Bartlett {\it et al.} 1995, Lineweaver this volume), 
or in a non-zero cosmological constant such 
that $\Omega{_o} = \Omega_{\sl matter} + \Omega_{\sl \Lambda}$, 
where 
$\Omega_{\sl \Lambda} \equiv{\Lambda/3} H_{o}^2$ but still, this is not 
consistent with dynamical evidence of large $\Omega_{o}$.
\begin{figure}[h]
\epsfxsize=6.5cm
\centerline{\epsfbox{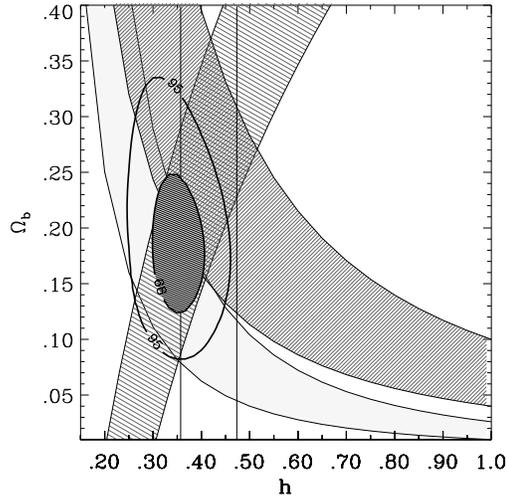}}
\caption{\small \sl This plot is from Lineweaver et al. (1997). It gives the region of the 
 h- $\Omega_b$ plane allowed by various constrains:BBN, cluster baryonic fraction (White et al.), and the shape parameter ${\Gamma}$ from galaxy and cluster scale density fluctuations. A low h value is preferred }
\end{figure}

\noindent Other possible solutions are:\\
\indent
1.- The calculations of standard primordial nucleosynthesis are incorrect.\\
\indent
2.- The X-ray gas is more concentrated with 
respect to the dark matter but White {\it et al.}(1993) have shown that gravitational and dissipative 
effects during cluster 
formation cannot account for such baryon overdensity.\\
\indent
3.- The intracluster gas is multi-phase, but no model has been proposed to explain 
such a clumpy configuration of the gas.\\
\indent
4.- There is a problem with the mass estimates. This solution will be 
discussed in next sections.\\

\noindent Finally, one may ask another interesting question. Is the baryon fraction the 
same in all clusters ? In the standard picture of cluster formation 
driven solely by gravitational instability and where cluster evolution is entirely self-similar, 
the expected baryon fraction should be constant, because no segregation 
between the gas and dark matter has occured. However, if we gather all the derived baryon 
fractions in the literature and compare them, the answer is clearly NO. For example the 
derived mean values of WF and DJF samples are different, {15\%} for the former and {20\%} for the later (but 
see Evrard 1997). More recently, Lowenstein \& Mushotzky (1996) have shown evidence of 
variations in 
baryon fraction from 
their analysis of two poor Abell clusters, A1060 and AWM 7, using the most 
recent X-ray observations 
from ROSAT and ASCA (Figure 2). Such variation in baryon fraction from cluster 
to cluster  
requires some process in addition to gravity, like feedback mechanisms or some other 
non-gravitational effects as suggested by DJF, but there are no theoretical 
arguments justifying such ideas (White {\it et al.} 1993).\\
\noindent
\begin{figure}[h]
\epsfxsize=7.5cm
\centerline{\epsfbox{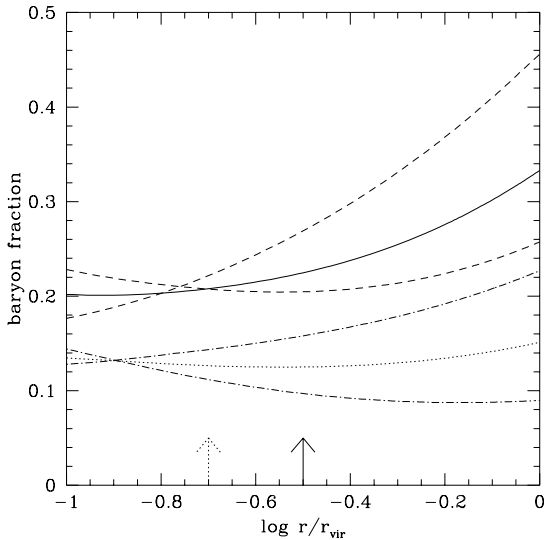}}
\caption{\small \sl This plot is taken from Lowenstein \& Mushotzky 
(1996). It gives the enclosed baryon fraction versus radius in units of the virial radius in 
A1060 and AWM7. The dotted (solid) line represents the best-fit mass
model,  the
dot-dashed (dashed) lines the most compact and diffuse models
for A1060 (AWM 7). The compact models lie below the best fits.}
\end{figure}
\noindent

\section{The dark matter problem}

Galaxies and hot gas are only a small part of the total cluster mass. The dominant component 
is the dark matter.
Zwicky in 1933 then Smith (1936) have shown that the virial mass exceeds by 
a large factor the luminous mass. This led them to 
invoke for the first time, the problem of missing mass. To quantify the 
amount of dark matter, we 
usually calculate 
the mass-to-light ratio (MLR). The mean value found for rich clusters using blue luminosties is
 $M/L_{B}\approx 300 {\sl h}$.  In this unit the $M/L{_B}$
required to 
close the Universe is $\sim 1200{\sl h}$. Therefore, if we assume that clusters of galaxies are 
good tracers of the 
whole Universe, then $\Omega_0 = 0.2 - 0.3$. So, if one believes in a matter dominated flat 
cosmological model ${\Omega_0} = 1$, then where is the missing mass? 
Still now we do not know how the dark matter is distributed relatively to the visible 
matter. Cluster mass determinations using optical observations are based on the 
assumption that mass follows the light distribution which is just an 
assumption not yet confirmed. 
What are the predictions on the mass 
distribution? Cosmological theories predict that dark matter is more diffusely 
distributed than galaxies. 
West and Richstone using N-body simulations have indeed confirmed this behaviour 
(West \& Richstone 1988).
Furtheremore Hughes (1989), using X-ray observations of the Coma cluster, has shown 
that models where 
dark matter parallels the distribution of hot gas are ruled out by the data. 
With the improvement of weak lensing analysis, one may hope that this question will be answered 
more precisely, in the near future. 

\section{Mass determinations}

In this section, I will review the methods used to estimate clusters masses and will discuss their validity.\\

\subsection{Optical methods}
\subsubsection{The Virial Theorem method}:

\noindent Early estimates of cluster masses (before X-ray observations became possible) were based on the 
application of the Virial theorem. If one assumes that clusters of galaxies are bound and self-gravitating 
systems then the {\sl virial} mass is given by:
 
\begin{equation}
M_{v} {\sim 3 \frac{R{_G}{\sigma}{_r}{^2}}{G}},
\end{equation}

\noindent where ${\sigma}^2$ and ${R_{G}}$ are evaluated from the radial velocity distribution (2) and the projected spatial distribution 
of a fair sample of galaxies.
As we have seen the naive application of equation(4) leads to large amounts of dark matter. 
Therefore the question we want to address here is: 
How secure are the virial mass estimates?
Projection effects, contamination by foreground galaxies and anisotropy of the velocity distribution 
may introduce uncertainties into the determination of the mass. But they are 
small effects 
and can not explain such large virial masses.    
Several observations at both optical and X-rays wavelengths provide convincing evidence of the presence 
of substructure in a large sample of clusters (Baier 1983, Bird, 1994, Mohr 
{\it et al.} 1993). X-ray imaging observations with the Einstein satellite first 
revealed such complex structure (Forman {\it et al.} 1981) in contrast to the 
smooth shape assumed in previous studies. Even clusters that exhibit a fairly 
smooth and apparently well-relaxed configuration , like the Coma cluster, have 
been found to contain substructure (Fitchett \& Webster 1987, Mellier {\it et al.} 1988) with a large subcluster 
centered on NGC 4839 that appears to be falling into the Coma cluster. If this subclustering is 
not correctly taken into account, this would introduce large uncertainties in the 
dynamical mass. 
A substructure with 10\% of the mass can introduce an underestimation of 40\% on 
the MLR.\\
However, the most serious problem of using the virial theorem comes from the fact that we do not know how 
the dark matter is distributed. Indeed, the application of the standard virial theorem assumes that mass
follows the light distribution. What happens when this 
assumption is relaxed? 
It has been shown (Sadat 1995) that in this case the standard application of the virial theorem introduces a bias 
on the cluster masses, and this bias (${\mu}=(M/L)_{\sl dyn}/(M/L)_{\sl true}$) depends strongly 
and in a non-linear 
way on the relative concentrations of the visible and invisible components. It is found that 
the cluster mass is over-(under) 
estimated if the dark matter is more (less) concentrated by an amount
\begin{equation}
{\mu}=\frac{{[1 + 2C{_c}R_{\sl true}+C_{\lambda}R_{\sl true}^2}]}{[1+C{_v}R_{\sl true}]},
\end{equation}
\noindent where $R_{\sl true}$ is the true ratio of the masses 
$M_{\sl DM}/M_{\sl gal}$ and $C_{v}$, $C_{\lambda}$, $C_{c}$ 
are the relative concentrations of the 2 components.
As an illustration of this  
effect we have plotted in Figure 3 the bias ${\mu}$ versus $R_{\sl true}$ in the case where 
the dark matter is less concentrated than the galaxies.
\noindent
\begin{figure}[h]
\centerline{\epsfbox{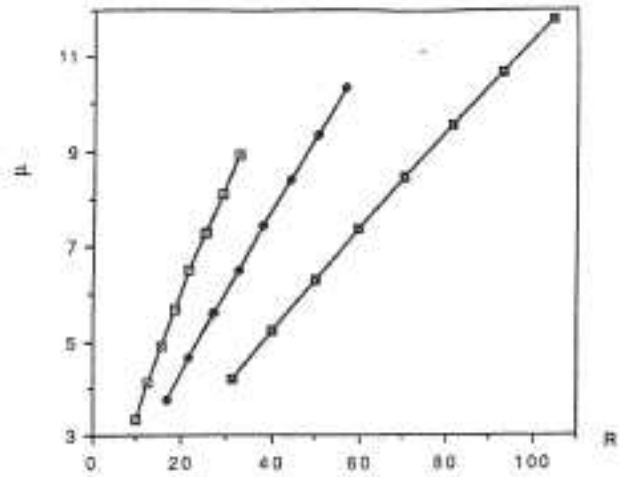}}
\caption{\small \sl The bias ${\mu}$ versus the true ratio ${R=R_{true}}$ of 
${dark/luminous}$ 
mass in the case where the mass is more diffusely distibuted than 
galaxies. The three curves correpond to different concentrations} 
\label{Fig. 3}
\end{figure}
\noindent
\begin{figure}[htp]

\centerline{\epsfbox{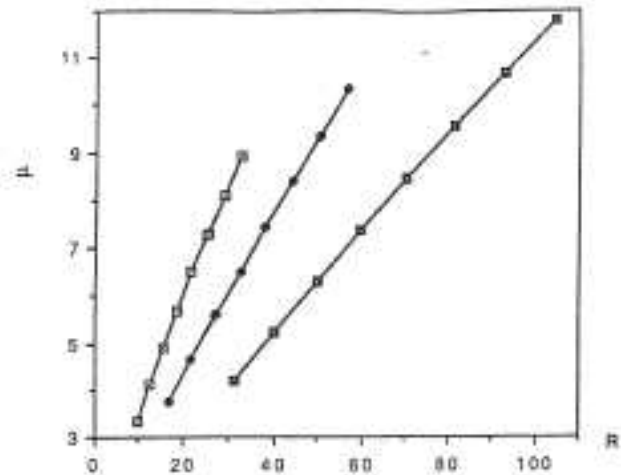}}
\caption{\small \sl Logarithm of the true ratio R  
versus the bias ${\mu}$ in the case where the mass is more concentrated than galaxies.}

\end{figure} 

\noindent Figure 3 shows that for a ratio of, say, $R_{\sl true} {\sim 30}$, the 
virial theorem leads to a dynamical ratio ${\sim 3}$ to 7 times lower. In this case the virial mass determination 
underestimates the 
true mass, while in the case where the dark 
matter is more 
concentrated (Figure 4) the dynamical ratio reaches the value of ${\mu}$ ${\sim 100}$ which is 
3 times higher than the true value ${\sim 30}$!. The true mass is overestimated.
Note, that in the mass- follows- light case, $C_{c} = C_{\lambda} = C_{v} = 1$ and
${\mu}$= 1 + $R_{\sl true}$. If one defines a new quantity $R_{dyn} = {\mu} - 1$ 
which measures the virial estimated ratio of 
dark matter mass to visible mass, one  can see that in this case 
$R_{dyn} = R_{\sl true}$ and therefore that the virial mass is 
equal to the true mass. 
It seems clear from this analysis, that as long as we do not know anything 
on the distribution of dark versus visible matter, one has to be ``sceptical'' about the  
masses derived from the ``virial'' method. \\
 
\subsubsection{Kinematic method}

If the system is in equilibrium, one can use the equation of stellar hydrodynamics to 
derive the mass \\
\begin{equation}
M(< r) = \frac{-G n_{\sl {gal}}(r)}{r^2} \left[\frac{dn_{\sl {gal}}(r)}{dr}{\sigma}_r(r){^2}+\frac{2 n_{\sl {gal}(r)}}{r}
[{\sigma}{_r}(r){^2}-{\sigma}{_t}(r){^2}]\right].
\end{equation}

\noindent For isotropic orbits, ${\sigma}_{r}$ =${\sigma}_{t}$, there would be a unique 
solution. Unfortunately the orbits of the galaxies are poorly known. Therefore, we have 
to solve this equation with three unknown quantities: ${\sigma}{_r}(r)$, ${\sigma}{_t}(r)$ 
and M(r). Generally, it is assumed that either ${\beta}(r) = 1 - {\sigma}{_t}(r)^2/{\sigma}{_r}(r)^2$ 
and M(r) are known functions and then derive $n_{\sl gal}$ and ${\sigma}{_r}(r)$ 
from eq. (6) which are consistent with observations. Unfortunately, the observed velocity 
dispersion profiles of clusters of 
galaxies are poorly known and can not put strong constraints on the mass. Indeed, 
even for the best studied Coma cluster, Merritt (1994) has 
shown that the observed velocity dispersion profile of this cluster 
is consistent with several mass distributions. 
 
\subsection{The hydrostatic isothermal ${\beta}$-model}

Problems encountered with optical methods like the shapes of galaxy orbits, the small 
number of galaxies in a cluster, effects of contamination and projection can be 
avoided by using the observations of the hot X-ray emitting gas. The gas can be treated as 
an isotropic fluid, since the elastic collision times for ions and electrons are much shorter than the 
timescales for cooling and heating. The timescale required for a sound wave in the intracluster gas to 
cross a cluster is given by 
$t{_X}=1.3 [T_{\sl gas}/10{^8}]^{-1/2} [R/1 Mpc]$ Gyr. \\
Furtheremore, since this time is shorter than the dynamical time of the cluster 
(${\sim 10}$ Gyr), the gas can be assumed to be in hydrostatic equilibrium with the cluster potential (Sarazin 1986). \\
Under the assumption of spherical symmetry the equation of hydrostatic equilibrium (balance between the 
pressure and the gravitational forces) can be solved for the mass interior to r, M(r):\\

\begin{equation}
M(r)=- \frac{kT_{gas}}{G{\mu}m{_p}}\left[\frac{dlog{\rho}_{gas}}{dlogr} +\frac{dlogT_{gas}}
{dlogr}\right],
\end{equation}
\noindent where $T_{gas}$(r) and ${\rho}_{gas}$(r) are the temperature and the gas density 
profiles, k
is Boltzmann's constant, and ${\mu}$ $m{_p}$ is the mean molecular weight of the gas.
In principle, the knowledge of $T_{gas}$ and ${\rho}_{gas}$ from the observations, directly yields the actual 
mass distribution M(r). This method has several advantages over the optical approach. The gas is 
isotropic, there are no contamination effects and the most important advantage is that 
the mass distribution is derived directly without any assumption about the dark matter 
distribution as is the case with the optical method.\\
The sad point is that one must recover three dimensional profiles from projected profiles.
For the temperature information, this requires the measurement of 
$T_{gas}$(r) which is still 
very difficult to obtain, even with the ASCA satellite. In practice, we assume that the gas is 
isothermal at a mean 
temperature ${T_{X}}$. Numerical simulations (Evrard 1996) and recent ASCA results (Ikebe {\it 
et al.} 1994) seem to support this 
assumption at least out to a radius of 1.5 Mpc. If the gas is isothermal,  
$\rho_{gas} \propto{n_{gal}^{\beta}}$, then the gas distribution is given by the 
following (Cavaliere \& Fusco-Femiano 1976)\\

\begin{equation}
\rho_{gas}= {\rho}_{gas}(0)[1 +(R/R_{c})^2]^{-3{\beta}/2},
\end{equation}

\noindent where $R_{c}$ is the core radius and ${\beta}$ is given by,
\begin{equation}
{\beta}=\frac{{\mu}m_{p}{\sigma}{^2}}{kT_{x}},
\end{equation} 
\noindent and ${\sigma}$ is the line of sight velocity dispersion. Both quantities 
are derived from the observed surface brightness profile which 
is found to be well characterized by a simple analytical form:\\
\begin{equation}
S(x)=S_o [1 + (x/R_{c})^2)]^{-3{\beta}_{fit}+1/2}
\end{equation}
\noindent This functional form gives relatively accurate fits to the data 
(Jones \& Forman 1984) except in the central regions of clusters where
cooling flows occur. Typical values of ${\beta}_{fit} {\sim}2/3$ are smaller 
than 
the value obtained using (9) from the measurements of ${T_x}$ and 
${\sigma}$. This 
discrepancy is the so called ${\it \sl {\beta} - problem}$ and has been 
thoroughly discussed in the litterature. 
Some solutions have been suggested to solve this problem (Bahcall \& Lubin 1994, 
Evrard 1990, Navarro {\it et al.} 1995) see also Gerbal {\it et al.} (1995). Smaller than 
this typical 
values are obtained by Durret {\it et al. 1995} with a mean value around 0.4. In their work, Durret {\it et al.} 
have analyzed a sample of 12 Einstein clusters with an improved method 
(Gerbal {\it et al. 1994}) of analysis which derives the density and 
temperature profiles of the X-ray gas by comparing a real cluster X-ray image to a "synthetic" image for which 
the counts predicted to be detected by the IPC was calculated by taking into account all the characteristics of 
the detector such as the point spread function, the effective area as a function of 
radius and energy. The ellipticity of the cluster is also taken into account. 
The resulting simulated images are fitted pixel per pixel to observed ones by minimizing the following function :\\

\begin{equation}
{\it {X^2}}=  {\sum}{\sum} \frac{(N_{\sl IPC}(b)-N_{\sl cts}(b))^2}{N_{\sl cts}(b)}
\end{equation}

\noindent A consequence of such flat (small ${\beta}$) gas density profiles is the derived gas mass to dynamical mass 
ratios (baryon fraction) which are exceedingly large.  Another interesting result of this analysis is the highly 
centrally peaked dark matter distribution in good agreement with the results based on the imaging and 
modelling of gravitational arcs in clusters (Tyson {\it et al.} 1990, 
Hammer 1991, Mellier {\it et al.} 1993, Wu \& Hammer 1993).\\
\medskip
Using isothermality and (8), equation (7) becomes :\\
\begin{equation}
M(r)= 3{\beta}/G \frac{kT_{X}r}{{\mu}m_{p}}\frac{(r/r_{c})^2}{1 + (r/r_c)^2}
\end{equation}
\noindent with ${{\mu}=0.59}$. This method has been extensively 
used to derive cluster masses, but still one 
may ask how secure this method is?\\
 The accuracy of the hydrostatic, 
isothermal ``beta-model'' method has been examined through hydrodynamical 
numerical simulations (Schindler {\it et al.} 1995, Evrard 1996). In particular, Evrard has 
shown that this method gives remarkably accurate masses inside a radius between 
0.5-2.5 Mpc but with a large scatter (15 - 30{\%}) (Figure 5).\\
\noindent
\begin{figure}[h]
\epsfxsize=15cm
\centerline{\epsfbox{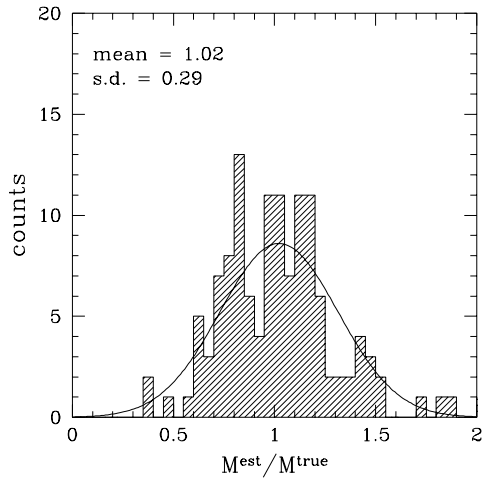}}
\caption{\small \sl Histograms of the estimated mass from the ${\beta}$ model from Evrard 1996}
\end{figure}
\noindent However, Bartelmann and Steinmetz (1996) 
have reached the opposite conclusion, they have found from their gas-dynamical simulations that the 
${\sl {\beta}- model}$ yields systematically low cluster mass estimates .\\ 
Furtheremore, Balland \& Blanchard (1995) have discussed the validity 
of using equation (7) to infer the mass M(r) from the observed temperature T(r).
They argue that the hydrostatic equilibrium equation is unstable and, 
using a Monte-Carlo 
procedure, that the resulting accuracy of the mass estimates is rather 
poor; larger than generally 
claimed. Applying their procedure to the Coma cluster, they find a factor of at 
least 2 uncertainty in the mass inside the Abell radius, even when the measurement of the 
temperature is improved using ROSAT data (Figure 6).
An alternative way to go round the ${\beta}-model$  i.e the surface brightness 
fitting is not required, has been suggested recently by Evrard (1996). This new method exploits an interesting result of his simulations, 
that is the tight relation between the mass and the temperature and uses the 
resulting scaling relations: $r_{500}(T_{X})\propto{T_{X}^{1/2}}$ and 
$M_{500}(T_{X}) \propto{T_{X}^{3/2}}$
\noindent which lead to more 
accurate masses and the scatter found in the ${\it \sl {\beta}-model}$ is then
eliminated (Figure 7). Of course such conclusions are given in the frame of numerical 
simulations which simulate clusters in ``somehow'' perfect conditions. For example their analysis 
uses the clusters emission- weighted temperature which comes from their 
simulations and not from cluster
spectra. 
Furthermore, the simulated X-rays images can be analyzed out to large radii which is not generally 
the case in real observed X-rays ones. Finally, the ${\it \sl {\beta}-model}$ 
method is based on the assumption of spherical 
symmetry. However, more often clusters exhibit a more complex morphologies due 
to the presence of substructures. Numerical simulations have 
demonstrated that masses of clusters which are undergoing a merging event, 
are generally under-estimated because part of the energy of the gas is in the kinetic form due to the bulk 
motion rather than in the thermal form, therefore the temperature of the gas is underestimated and so are the 
clusters masses. The underestimation of the mass due to the presence of substructure can reach 
{40\%} (Schindler 1996).
\clearpage
\noindent
\begin{figure}[h]
\epsfxsize=6.5cm
\centerline{\epsfbox{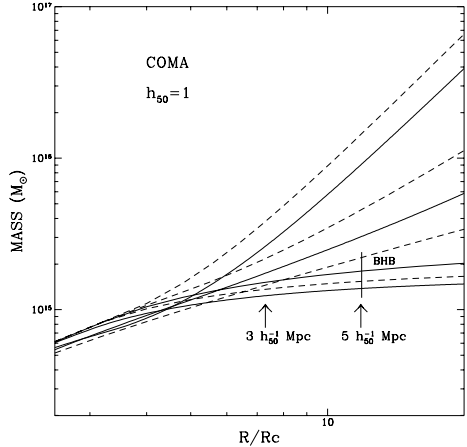}}
\caption{\small \sl This plot shows how the mass profile is poorly constrained by 
observed temperature profiles even with the recent results on Coma cluster from ROSAT 
(Briel et al. 1992) }
\end{figure}
\noindent
\begin{figure}[t]
\epsfxsize=15cm
\centerline{\epsfbox{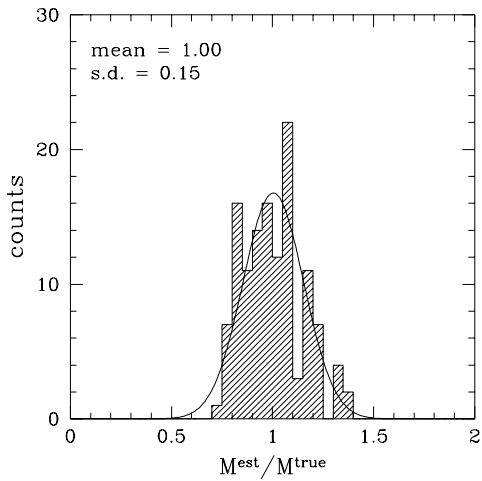}}
\caption{\small \sl  Histograms of the estimated mass from 
the scaling law from Evrard 1996}
\label{Fig. 7}
\end{figure}
\noindent
\subsection{The Gravitational lensing mass estimates}
The discovery of giant blue luminous arcs in clusters A370 and Cl 2244-02 
(Soucail {\it et al.} 1987, Lynds \& Petrosian 1989) has provided the first observational 
evidence that clusters of galaxies may act as gravitational lenses on background 
galaxies, a 
possibility which was first discussed by Noonan (1971). 
Gravitational lensing provides a very powerful tool to directly 
measure the projected mass 
distribution. This method, presents many advantages over the X-ray mass estimates, for example,  it does not require any assumption 
on the mass distribution or on the dynamical state of the cluster.
Since the pionnering work by Tyson {\it et al.} (1990). It has become more and more common 
to use weak gravitational lensing to 
map the dark matter distribution in clusters.
Detailed study of image formation through gravitational lensing can be found in the review by 
Schneider, Ehlers and Falco ( 1992) and Fort \& Mellier (1994). I will 
just summary very briefly 
the manifestations of the lensing effect and the way the lensing masses are derived. 
The lensing effects can be divided into two main regimes depending on the lens
 configuration :\\

\indent
1-{\sl  The strong lensing regime} :\\

\noindent The distorsion of distant galaxies by foreground  clusters of galaxies gives rise to the 
spectacular strong 
arcs observed in the central regions of clusters e.g. A370 
corresponding to a large magnification and strong distorsion.  
The arclet regime is intermediate between the arc and the weak distorsion 
regimes.\\
\indent
2-{\sl The weak lensing regime or weak shear}:\\

\noindent The first observational detection using optical galaxies as sources is due to 
Tyson, Valdes and Wenk (1990). In this case, each 
source produces only one image which experiences only a weak distorion of its shape. \\
\noindent
The strong lensing regime constrains the total mass enclosed within the ``
Einstein radius'', while weak shear 
effects determine the distribution of the mass at the outer regions (see Brainherd in these proceedings).

\subsubsection{Constraints from Strong Lensing}
The projected cluster mass within the Einstein radius $r_{E}$ of arc or arclet can be easily derived if one assumes a 
spherical matter distribution for the lensing cluster and assuming that the system 
observer-lens-source is aligned along the line of sight\\
\begin{equation}
M_{\sl lens} = {\pi}r_{E}^2 {\sum}_{\sl crit} ,
\end{equation}

\noindent where ${\sum}_{\sl lens}= \frac{c^2}{4 {\pi} G}\frac{D_{s}}{D_{\sl l}D_{\sl ls}}$ is the 
critical mass density with 
${D_{s}}$, ${D_{\sl l}}$ and ${D_{\sl ls}}$ being the distance to the source (the galaxy), 
distance between the source and the 
lensing cluster and the distance between the source and the cluster respectively.
For more complex configurations, cluster masses are estimated by lens modelling 
(see Fort \& Mellier 1994 for a review). This method, however gives the mass inside 
the radius
where the arcs are observed which are usually very small ${\approx 50}$ kpc.\\ 

\subsubsection{Constraints from weak lensing}
To construct the surface mass density profile one uses the statistic suggested by 
(Fahlman {\it et al.} 1994) 
\indent
\begin{equation}
{\chi} (r_{1}, r_{2})=\overline{\sum}( r_1)-\overline{\sum}{(r_1 < r < r_2)}=
\frac{2\sum_{\sl crit}}{1 - r^{2}_1/r^{2}_2} \int_{r_1}^{r_2}<{\epsilon}> dlnr 
\end{equation}
\noindent where $<{\epsilon}>$ is the mean tangential component of the image ellipticities.  This method has been 
successfully applied to several clusters (see Table 1).
\noindent
How reliable is the lensing method? The main shortcomings of the lensing method is that the application of 
equation (14) requires an estimation of ${\sum}_{\sl crit}$ and therefore the knowledge of the 
redshift of the sources 
which is difficult to obtain. This may introduce large uncertainty in the mass especially 
for distant clusters. On the other hand it is not possible to obtain a true value of the mass only from the shear map, 
even in the best case where the sources redshift is known because of the degeneracy due to 
the fact that the addition of a constant mass plane does not induce any shear on background 
galaxies. This degeneracy may be broken by measuring the magnification ${\mu}$ of the background 
which gives an absolute measurement of the mass. Broadhurst {\it et al.} (1995) have proposed a very nice 
method to measure ${\mu}$ by comparing the number count in a lensed and unlensed field. They 
find that depending on the slope of the number count in the reference field s=dlogN(m)/dm, 
they observe more or 
fewer objects in the lensed field. In the case where blue galaxies are selected, the 
counts are 
unaltered, since the slope is in this case equal to the critical value s=0.4. This method has been 
applied successfully to the cluster A1689 by Broadhurst (1995). The weakness of this 
method is that it requires the measurement of the shape, size and magnitude of very faint 
objects. Van Waerbeke {\it et al.} (1996) have recently suggested a new method 
to analyz<e 
the lensing effects which avoids the measurement of the shape parameter. 
But still, the weak lensing method leads to very encouraging results and 
promises to yield unambigeous 
information about the mass distribution in the near future.
\subsection{Comparison between X-ray and lensing cluster mass estimates}
Miralda-Escud\'e \& Babul (1995) have raised an interesting puzzle. They found from their analysis of  
Abell clusters A2218, A1689
that the mass in the central part of the cluster inferred from the strong lensing method is greater than
that derived from the X - ray method by a factor of  2 - 2.5. Wu \& Fang have gathered all the clusters 
for which the mass has been estimated and compared the X-ray to lensing masses. They have found a 
systematic discrepancy between the two masses at small radius 
${\approx 0.25}$ ${h_{50}^{-1}}$ {\rm Mpc} 
which vanishes at larger radii. However the lensing and the X-ray information in their 
sample do not come from the same cluster. Early studies based on 
both optical and lensing observations have led to the same conclusion: 
there is a cluster mass discrepancy by the same factor (Wu \ Fang 1994, Fahlman 
{\it et al.} 1994). But, it seems from a recent statistical analysis that 
{\sl virial} masses are 
consistent with gravitational lensing masses (Wu \& Fang 1997).
The disagreement between the lensing masses and X-ray masses may be due to the fact that X-ray analysis, namely the 
${\sl {\beta}-model}$, underestimates the masses. Indeed, the assumption of 
hydrostatic equation may be invalid, because of several reasons, non-thermal 
pressure, merging effects, a multi-phase medium, unstability of the 
equilibrium equation etc...Unfortunately,  it is hard to quantify all these effects and to know 
which is the most important one.
\begin{table}[h]
\scriptsize
\begin{center}
\begin{tabular}{lllllcllclllcr}
\tableline
  &  &  &  &  &  &  \\
cluster & $Redshift$ & arc/w.l.&  $r$(Mpc) &   
$m_{lens}$  & ref. \\ 
\tableline
A370 & 0.374 & arc  & 0.16/0.4 & 2.9/12. &
         1\\
A1689 & 0.17 & arc/w.l. &0.19/3. & 3.6/89. &
         2,3\\
A2163 & 0.201 & arc/w.l. & 0.066/0.9 & 0.41/$13_{-7}^{+7}$ & 
         4,5\\
A2218 & 0.175 & arc/w.l. & 0.085/0.8 & 0.61/7.8&
         6\\
A2219 & 0.225 & arc & 0.1&   1.6 &
         7\\
A2390 & 0.231 & arc/w.l. &0.18/1.15 & 1.6/$19.5_{-6.5}^{+6.5}$ &
         8,9 \\
CL0500 & 0.316 & arc & 0.15 & 1.9 & 
         10\\
CL0024 & 0.391 E & arc/w.l & 0.22/3.0 & 3.6/40. & 
         11,12\\
CL0302 & 0.423 & arc & 0.12&  1.6 &
         13\\
CL2244 & 0.328 & arc & 0.06&   0.25 & 
         14\\
MS1224 & 0.33 & w.l. &  0.96 & 7.0 & 
         15\\
MS1054 & 0.83 & w.l. & 1.9 & $28_{-6}^{+6}$ & 
         16 \\
AC114 & 0.31 & arc &0.35 & 13 & 
         17\\
PKS0745 & 0.103 & arc & 0.046 & 0.30 &
         18\\
RXJ1347 & 0.451 & arc & 0.24 & 6.6 &
         19\\
\tableline
\end{tabular}
\end{center}
\caption{\small Lensing masses for a sample of clusters.
${\bf Ref.}$  (1)Kneib et al. 1993; (2,3) Tyson \& Fisher 1995; (4,5) Miralda-E \& Babul 1995, 
Squires et al. 1996a; (6) Kneib et al. 1996,(7) Smail et al. 
1995b; (8,9) Pello et al. 1991, Squires et al. 1996b;(10) 
Giraud 1988; (11, 12) Wallington \& Kochanek 1995, Bonnet et al. 1994;
(13)Mathez et al. 1992,(14) Hammer et al. 1989, (15) Fahlman et al. 
1994; (16) Luppino \& Kaiser 1996;(17)Smail et al. 1995a; (18) 
Allen et al. 1996; (19)Schindler et al. 1995}
\label{table-1}
\end{table} 

\section{Discussion and conclusion}

Dynamical analysis of clusters of galaxies have led to two important results: 
the presence of large amount of dark matter and the evidence of high baryonic fraction, both 
have implications on cosmology through ${\Omega}_{0}$ and ${\Omega}_b$, the 
density of the Universe and its baryon content respectively. Estimating the masses 
of clusters of galaxies, is not straightforward, because it depends on the 
validity of the assumptions underlying the method from which the mass is 
determined, the {\it mass-follows-light} in the case of the {\sl virial} masses, 
the hydrostatic equilibrium and isothermality of the gas for the X-ray mass 
determination. Gravitational lensing methods provide with a new strong tool to 
constrain both the amount of mass and its distribution. Comparing the X-rays to 
lensing masses give rise, at least in the inner part 
of the cluster, to the mass discrepancy problem. The most probable explanation, 
would be the underestimation of the X-ray mass. The interesting implication, 
is that clusters would be more massive than we think, and the ratio of 
gas mass to total mass (the fraction of baryons) could be in more better 
agreement with nucleosynthesis predictions and an ${\Omega}_0$=1 Universe. 
Finally, thanks to new recent set of observations, it appears that {\sl virial}
masses are in good agreement with the lensing masses (Wu \& Fang 1997), if this result is 
true, that means that the virial masses are accurate and one may conclude that 
indeed, the mass follows light, since it is 
only in this case that the {\sl virial} method gives accurate mass determination 
(Sadat 1995).\\
\\
\medskip
\acknowledgments
First I would like to thank Drs. K. Chamcham, M. Henry and D. Valls-Gabaud for 
the invitation to the School.
I'm grateful to C. Balland, C. Lineweaver, X.P. Wu, and A. Evrard, for providing the postscript files 
of their figures (I apologize for the rather poor quality of the few 
scanned figures). Finaly, I would like to thank our morrocan hosts for their 
hospitality.
This work 
was supported by the French {\it Minist\`ere National de l'Enseignement Sup\'erieur et de 
la Recherche}.\\

\end{document}